# Passive damping of beam vibrations through distributed electric networks and piezoelectric transducers: prototype design and experimental validation

**F dell'Isola**[1], **C Maurini**[2,3] **and M Porfiri**[1,4]

[1] Dipartimento di Ingegneria Strutturale e Geotecnica, Università di Roma La Sapienza, via Eudossiana 18, 00184, Roma, Italy
[2] Dipartimento di Meccanica e Aeronautica, Università di Roma La Sapienza, via Eudossiana 18, 00184, Roma, Italy
[3] Laboratoire d'Etudes Méchaniques des Assemblages (FRE 2481), Université de Versailles/Saint-Quentin-en-Yvelines, 45 Ave des Etats-Unis, 78035 Versailles, France
[4] Department of Engineering Science and Mechanics, Virgina Polytechnic Institute and State University, Blacksburg, VA 24061, USA

**Abstract**
The aim of this work is two-fold: to design devices for passive electric damping of structural vibrations by distributed piezoelectric transducers and electric networks, and to experimentally validate the effectiveness of such a damping concept. Two different electric networks are employed, namely a purely resistive network and an inductive–resistive one. The presented devices can be considered as distributed versions of the well-known resistive and resonant shunt of a single piezoelectric transducer. The technical feasibility and damping effectiveness of the proposed novel devices are assessed through the construction of an experimental prototype. Experimental results are shown to be in very good agreement with theoretical predictions. It is proved that the presented technique allows for a substantial reduction in the inductances used when compared with those required by the single resonant shunted transducer. In particular, it is shown that the required inductance decreases when the number of piezoelectric elements is increased. The electric networks are optimized in order to reduce forced vibrations close to the first resonance frequency. Nevertheless, the damping effectiveness for higher modes is experimentally proved. As well as specific results, fundamental theoretical and experimental considerations for passive distributed vibration control are provided.

(Some figures in this article are in colour only in the electronic version)

## 1. Introduction

Piezoelectric elements together with passive electrical networks can be used to damp structural vibrations (see e.g. [1] and [2]). In [3] it was proposed that mechanical energy can be dissipated by shunting a single piezoelectric transducer with either a simple resistor ($R$ impedance), or a resistor connected with an inductor ($RL$ impedance). Resistive shunting parallels viscoelastic damping: its effectiveness is wide-band, its technical implementation is immediate, but its performance is limited. On the other hand, a piezoelectric element shunted with a resistor and an inductor parallels a damped vibration absorber: its performance is impressive, but its effectiveness is narrow-band and its technical implementation is involved. The efficiency of the latter approach is guaranteed by the resonant coupling between the mechanical structural mode to be damped and the electric circuit constituted by the shunting $RL$ impedance and the inherent capacitance of the piezoelectric transducer. However, for typical values of inherent piezoelectric capacitances (10–100 nF), very high inductances (10–1000 H) are needed to tune the electrical resonance frequency to the structural one. Furthermore, the large internal parasitic dissipation of such a large inductor may exceed the optimal design dissipation for suppression of low-frequency vibration.

In order to overcome some of these drawbacks, some authors have proposed (see e.g. [4] and [5]) placing an additional capacitance across the transducer terminals thus reducing the optimal shunting inductance. Nevertheless, as underlined also in [5], an increase in the overall capacitance (inherent piezoelectric and additional one) induces a loss of performance (see e.g. the experimental results in figures 7 and 8 of [4]).

An alternative approach to the damping of structural vibrations exploits distributed piezoelectric elements and passive electric impedances to form a so-called piezoelectromechanical (PEM) structure. This technique was first proposed in [7], where an array of piezoelectric elements was positioned on a host beam, and the electric terminals of each adjacent pair interconnected via a floating inductor. By assuming a continuum model for the overall modular electromechanical system, it was theoretically shown that the maximum coupling effectiveness is achieved by tuning the electric circuit to a given mechanical mode. The optimal line inductance is reduced when the number of piezoelectric elements is increased (with respect to the number necessary to separately shunt each piezoelectric element, according to [3]). Further developments of the concept of passive distributed control and applications to multimodal damping are summarized in [6]. As the main result, it was theoretically shown that when the interconnecting impedances together with the inherent piezoelectric capacitances form a passive electric network duplicating the modal properties of the beam, efficient multimodal vibration damping is ensured. In [8], a comparison of the damping performances of distributed electric networks was made by studying electromechanical wave propagation in PEM beams.

All the cited works on passive electric distributed control were based on simplified models, regarding the electric network as a continuous medium and uniformly spreading the piezoelectric coupling on the structure. Moreover, the mentioned results have not been supported by experimental evidence, and the challenging problems implied in building a prototype have not been addressed.

In the present paper we design a *PEM beam prototype* obtained by interconnecting distributed piezoelectric elements either by $RL$ impedances ($RL$ network), or $R$ impedances ($R$ network) as in figure 1. This prototype is used to prove the technical feasibility of PEM structures and the effectiveness of the underlying concept of distributed control in reducing tuning inductances without affecting damping performance.

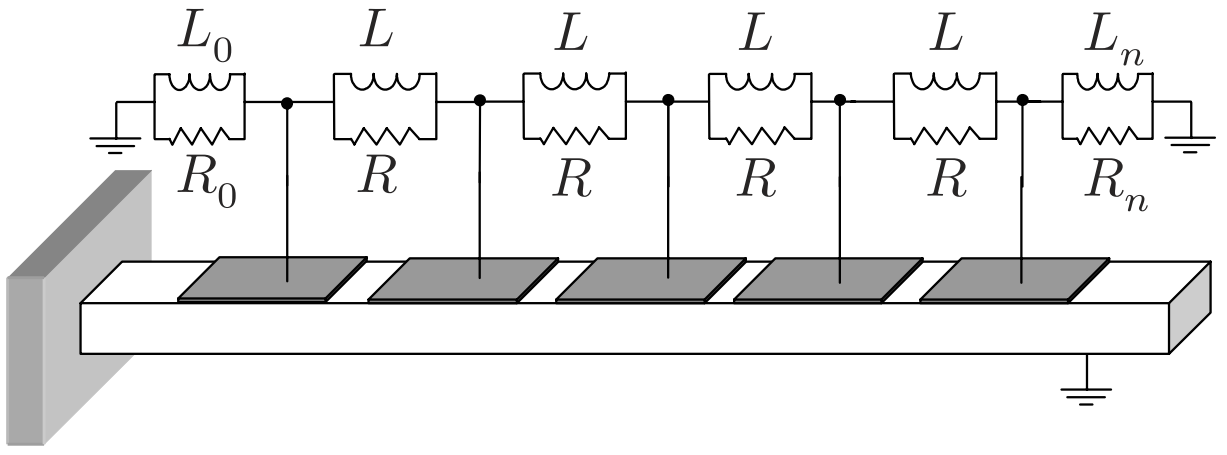

**Figure 1.** Sketch of a PEM beam with $RL$ impedances.

When designing an experimental prototype, the continuum model proposed in [7] does not seem to be sufficiently reliable: therefore, a refined model accounting for material discontinuities and the lumped nature of electric networks is adopted. The modelling approach is presented in section 2. In section 3, optimal line inductance and resistance are derived, and optimal electric boundary conditions are found. In section 4, the experimental set-up is described and the main results are presented. Section 5 is devoted to discussions and conclusions.

## 2. Mathematical modelling

A beam hosting a uniformly distributed array of electrically interconnected piezoelectric transducers is studied (see figure 1). An $RL$ network is obtained when each pair of adjacent transducers is connected via a floating $RL$ impedance; a simple resistive network is obtained by discarding the inductance. A sufficiently refined model of the device consists of an Euler beam piezoelectrically coupled to the introduced lumped interconnecting network completed by the inherent piezoelectric capacitances (see section 2.1). With the aim of controlling the vibration of the first structural mode, the electric network is designed by means of a simplified model describing only the coupling between the first electrical and mechanical modes (see section 2.2).

### 2.1. Refined model

By assuming a simple 1D Euler–Bernoulli-like model for both the beam and the attached piezoelectric elements, the state of the PEM beam in figure 1 is described by the scalar field $w$, representing the transverse deflection of the beam axis, and the vector $\{\psi_i\}_{i=1}^{n}$ of nodal flux linkages at the terminals of the $n$ piezoelectric transducers (the flux linkage $\psi_i$ is the time integral of the nodal voltage). Therefore, the corresponding evolution equations, when a mechanical excitation $f(x,t)$ is applied (and where $x$ and $t$ denote space and time variables, and the primes and superimposed dot indicate space and time distributional derivatives) are:

$$
\begin{aligned}
&(k(x)w''(x,t))'' + \sum_{i=1}^{n} g_i \dot{\psi}_i(t) p_i''(x) \\
&\qquad + \rho(x)\ddot{w}(x,t) = f(x,t), \\
&\sum_{j=1}^{n} Y_{ij}^{L}\psi_j(t) + \sum_{j=1}^{n} Y_{ij}^{R}\dot{\psi}_j(t) \\
&\qquad + c\chi_i\ddot{\psi}_i(t) - g_i\int_0^l \dot{w}(x,t)p_i''(x)\,\mathrm{d}x = 0.
\end{aligned}
\tag{1}
$$

Here $p_i(x)$ is equal to 1 in the region of the beam covered by the $i$th transducer and vanishes elsewhere; the bending stiffness $k(x)$ and the mass per unit length $\rho(x)$ are the sum of the beam and transducers contributions, namely

$$k(x) = k^{\text{beam}} + k^{\text{piezo}} \sum_{i=1}^{n} p_i(x),$$
$$\rho(x) = \rho^{\text{beam}} + \rho^{\text{piezo}} \sum_{i=1}^{n} p_i(x), \tag{2}$$

where $g_i$ denotes the piezoelectric coupling coefficient, $c_i = c\chi_i$ indicates the inherent capacitance of the $i$th piezoelectric transducer, $c$ is the average of the transducer capacitances, and the *tridiagonal* electric admittance matrices of the interconnecting network are given by:

$$[Y_{ij}^R] = \frac{1}{R}[\mathcal{N}_{ij}], \qquad [Y_{ij}^L] = \frac{1}{L}[\mathcal{N}_{ij}], \tag{3}$$

with

$$[\mathcal{N}_{ij}] = \begin{bmatrix} \frac{2}{1+\alpha_0} & -1 & 0 & \cdots & 0 \\ -1 & 2 & -1 & \cdots & \cdots \\ 0 & -1 & \cdots & \cdots & 0 \\ \cdots & \cdots & \cdots & 2 & -1 \\ 0 & \cdots & 0 & -1 & \frac{2}{1+\alpha_n} \end{bmatrix}. \tag{4}$$

The floating inductances interconnecting adjacent transducers are assumed to all be equal ($L$), while the boundary grounded inductances $L_0 = \frac{1+\alpha_0}{1-\alpha_0}L$ and $L_n = \frac{1+\alpha_n}{1-\alpha_n}L$ are left as free parameters; the chosen electric dissipation introduces only proportional damping (consequently the boundary resistances are $R_0 = \frac{1+\alpha_0}{1-\alpha_0}R$ and $R_n = \frac{1+\alpha_n}{1-\alpha_n}R$). The non-dimensional coefficients $(\alpha_0, \alpha_n)$ have been defined in order to span all the possible values for the boundary impedances for $(\alpha_0, \alpha_n) \in [-1, 1) \times [-1, 1)$. The non-uniform characteristics of the transducers along the array are accounted for by introducing the $\chi_i$ coefficient in equation (1), thus modelling the unavoidable imperfections of real prototypes.

If the mechanical excitation is provided through an additional piezoelectric actuator driven by a voltage $\dot{\psi}_{\text{a}}$ (the subscript a denotes quantities referring to the actuator), the forcing term $f$ becomes

$$f = g_{\text{a}}\dot{\psi}_{\text{a}}(t)p_a''(x), \tag{5}$$

and the stiffness and mass per unit length are modified by adding the corresponding contributions.

### 2.2. Reduced model

Let $\tilde{w}^{(i)}(x)$ and $\omega_i$ be the $i$th mechanical modal shape and angular frequency of the PEM beam with short-circuited electric terminals, i.e. for $\{\psi_i = 0\}_{i=1}^n$. Moreover let $\{\tilde{\psi}_j^{(h)}\}$ and $\lambda_h$ be the $h$th eigenvector and eigenvalue of the electric circuit constituted by the blocked PEM beam (i.e. a PEM beam with $w(x, t) = 0$). Therefore we define the mechanical modal shape and angular frequency so that, for cantilever beam boundary conditions,

$$(k(x)(\tilde{w}^{(i)}(x))'')'' = \omega_i^2 \rho(x)\tilde{w}^{(i)}(x),$$
$$\frac{1}{m_{\text{tot}}}\int_0^l \rho(x)\tilde{w}^{(i)}(x)\tilde{w}^{(j)}(x)\,\mathrm{d}x = \delta_{ij} \tag{6}$$

where $\delta_{ij}$ is the Kronecker delta and $m_{\text{tot}}$ is the beam mass. The $h$th electric eigenvector and eigenvalue are defined by

$$\sum_{j=1}^{n} \mathcal{N}_{ij}\tilde{\psi}_j^{(h)} = \lambda_h \chi_i \tilde{\psi}_i^{(h)}, \qquad \sum_{j=1}^{n} \tilde{\psi}_j^{(h)}\chi_j\tilde{\psi}_j^{(k)} = \delta_{hk}. \tag{7}$$

If the mechanical deflection is approximated by its expansion on the first $N$ modal shapes, by assuming

$$w(x, t) \simeq \sum_{i=1}^{N} \tilde{w}^{(i)}(x)W_i(t) \tag{8}$$

the following coupled evolution equations for the modal coefficients $\{W_i(t)\}_{i=1}^N$ and the flux linkages $\{\psi_h(t)\}_{h=1}^n$ are obtained from equations (1):

$$\ddot{W}_i(t) + \omega_i^2 W_i(t) + \frac{1}{m_{\text{tot}}}\sum_{j=1}^{n} G_{ij}\dot{\psi}_j(t) = F_i(t),$$
$$\frac{1}{c}\sum_{j=1}^{n} Y_{ij}^L \psi_j(t) + \frac{1}{c}\sum_{j=1}^{n} Y_{ij}^R \dot{\psi}_j(t) + \chi_j\ddot{\psi}_j(t) \tag{9}$$
$$-\frac{1}{c}\sum_{i=1}^{N} G_{ij}\dot{W}_i(t) = 0,$$

where the coupling and forcing modal parameters are defined by:

$$G_{ij} = g_j \int_0^l p_j(x)(\tilde{w}^{(i)}(x))''\,\mathrm{d}x,$$
$$F_i(t) = \frac{1}{m_{\text{tot}}}\int_0^l f(x, t)\tilde{w}^{(i)}(x)\,\mathrm{d}x. \tag{10}$$

By introducing the non-dimensional variables

$$W_i^* = \frac{W_i}{W_0}, \qquad \psi_i^* = \psi_i\sqrt{\frac{m_{\text{tot}}}{c}}W_0,$$
$$t^* = \omega_1 t, \qquad F_i^*(t) = \frac{F_i(t)}{W_0\omega_1^2}, \tag{11}$$

the following non-dimensional form for the governing equations is found:

$$\ddot{W}_i^*(t) + \frac{\omega_i^2}{\omega_1^2}W_i^*(t) + \sum_{k=1}^{n}\gamma_{ik}\dot{\psi}_k^*(t) = F_i^*(t),$$
$$\frac{1}{\omega_1^2 Lc}\sum_{k=1}^{n}\mathcal{N}_{jk}\psi_k^*(t) + \frac{1}{Rc\omega_1}\sum_{k=1}^{n}\mathcal{N}_{jk}\dot{\psi}_k^*(t) \tag{12}$$
$$+\ \chi_j\ddot{\psi}_j^*(t) - \sum_{i=1}^{N}\gamma_{ij}\dot{W}_i^*(t) = 0,$$

where

$$\gamma_{ij} = \frac{G_{ij}}{\omega_1\sqrt{c}}. \tag{13}$$

If the frequency range of the mechanical excitation and the first electrical resonance both lie in the neighbourhood of the first mechanical natural frequency, the coupling between the first mechanical and electric mode is dominant, and the influence of higher modes remains negligible. Hence it is reasonable to assume two degrees of freedom for the PEM beam, by assuming

$$w(x, t) \simeq \tilde{w}^{(1)}(x)W_1(t), \qquad \psi_j(t) \simeq \tilde{\psi}_j^{(1)}\Psi_1(t), \tag{14}$$

where $\Psi_1(t)$ is the first electrical modal coefficient. The corresponding non-dimensional evolution equations are

$$\begin{aligned} \ddot{W}_1^*(t) + W_1^*(t) + \gamma\dot{\Psi}_1^*(t) &= F_1^*(t) \\ \ddot{\Psi}_1^*(t) + \delta\dot{\Psi}_1^*(t) + \beta\Psi_1^*(t) - \gamma\dot{W}_1^*(t) &= 0, \end{aligned} \quad (15)$$

where the relevant non-dimensional parameters are defined as

$$\beta = \frac{\lambda_1}{Lc}\frac{1}{\omega_1^2}, \qquad \delta = \frac{\lambda_1}{Rc}\frac{1}{\omega_1}, \qquad \gamma = \sum_{j=1}^{n}\gamma_{1j}\tilde{\psi}_j^{(1)}. \quad (16)$$

The parameter $\beta$ measures the electromechanical tuning, since it is nothing more than the square of the ratio between the mechanical and the electrical frequencies. On the other hand, the parameter $\delta$ represents the modal damping of the electric circuit.

## 3. Prototype design: electrical system optimization

In this section we examine the damping performance of two different electric interconnections, namely:

(1) A resistive–inductive circuit ($RL$ network), obtained by interconnecting each pair of adjacent transducers via a floating $RL$ impedance, and shunting boundary elements by grounded $RL$ impedances.
(2) A resistive circuit ($R$ network), which is obtained from the resistive–inductive one by removing the inductors.

The former system is a distributed version of a resonant shunted piezoelectric transducer (see e.g. [3]), which is an electric version of a damped vibration absorber (see [9]); the latter is a distributed version of a resistive shunted piezoelectric transducer (see e.g. [3]), which is the electric counterpart of a viscoelastic damper (see [9]).

### 3.1. $RL$ network

The optimal design of the $RL$ network is obtained by the following steps:

(1) choosing a performance index,
(2) finding optimal non-dimensional tuning $\beta$ and dissipation $\delta$ parameters,
(3) finding the optimal boundary conditions and optimal impedances ($R$, $R_0$, $R_n$, $L$, $L_0$, $L_n$).

A reasonable and widespread (see e.g. [3]) criterion for damping optimization in the frequency domain is based on the $\mathcal{H}_\infty$ minimization of the mechanical mobility function, say $H_{RL}(\omega)$, obtained by manipulating the Fourier transform of equations (15):

$$\begin{aligned} H_{RL}(\omega) &:= \frac{\mathcal{F}[\dot{W}_1^*](\omega)}{\mathcal{F}[F_1^*](\omega)} \\ &= -\frac{\mathrm{j}\omega(-\omega^2+\beta+\mathrm{j}\omega\delta)}{-\omega^4+\omega^3\mathrm{j}\delta+\omega^2(\beta+1+\gamma^2)-\mathrm{j}\omega\delta-\beta} \end{aligned} \quad (17)$$

$\omega$, being the dimensionless angular frequency and $\mathcal{F}$ the Fourier transform. The $\mathcal{H}_\infty$ optimization consists in minimizing the maximum amplitude of the mechanical mobility function.

#### 3.1.1. *Optimal tuning and damping coefficients.*

As suggested in [9], the optimization problem is solved by exploiting some peculiar properties of the transfer function. In particular, it can be prove that there exist two points (*fixed points*) $S = (\omega_S, |H_{RL}(\omega_S)|)$ and $T = (\omega_T, |H_{RL}(\omega_T)|)$, which are independent of the parameter $\delta$. The optimal tuning parameter $\beta$ is found by imposing the transfer function to attain the same amplitude at the two frequencies $\omega_S$ and $\omega_T$, i.e.

$$|H_{RL}(\omega_S)| = |H_{RL}(\omega_T)|. \quad (18)$$

Furthermore, the dissipation parameter $\delta$ is found by imposing the *matched* transfer function to satisfy:

$$\frac{\mathrm{d}}{\mathrm{d}\omega}|H_{RL}(\omega)|\bigg|_{\omega=\omega_T} = \frac{\mathrm{d}}{\mathrm{d}\omega}|H_{RL}(\omega)|\bigg|_{\omega=\omega_S} = 0, \quad (19)$$

i.e. to obtain a flat transfer function in the neighbourhood of the mechanical resonance. The following values must be set for the non-dimensional parameters in order to fulfil all the given conditions:

$$\beta = 1, \qquad \delta = \sqrt{2/3}\gamma. \quad (20)$$

Equation (20) shows in particular that the maximum damping effectiveness is achieved when the electric network is tuned so as to resonate at the natural mechanical frequency. The following optimal line inductance $L$ and resistance $R$ are found by substituting the tuning conditions (20)

$$L_{\text{opt}} = \frac{\lambda_1}{(\omega_1)^2 c}, \qquad R_{\text{opt}} = \frac{\lambda_1}{\omega_1 c\sqrt{2/3}\gamma}. \quad (21)$$

Moreover, when $\beta = 1$

$$\omega_T - \omega_S = \frac{\gamma}{\sqrt{2}}. \quad (22)$$

Equation (22) can be used to accurately estimate the dimensionless coupling coefficient appearing in (15). The $\mathcal{H}_\infty$ norm of the transfer function when conditions (20) are satisfied is given by:

$$\|H_{RL}\|_\infty = |H_{RL}(\omega_S)| = |H_{RL}(\omega_T)| = \frac{\sqrt{2}}{\gamma}. \quad (23)$$

The above equation shows that the effectiveness of the vibration damping is proportional to the modal coupling $\gamma$.

#### 3.1.2. *Optimal boundary conditions.*

The dependence of performance on the modal coupling (see equation (23)) shows that once conditions (20) have been fulfilled, the electric network should be designed so as to maximize $\gamma$. Thus, the following auxiliary optimization problem is stated with the aim of choosing proper boundary impedances: *find* $\alpha_0$, $\alpha_n$ *to maximize*

$$\gamma(\alpha_0, \alpha_n) = \sum_{j=1}^{n}\gamma_j\tilde{\psi}_j^{(1)}(\alpha_0, \alpha_n), \quad (24)$$

where the explicit dependence of the electric modal shape on the boundary impedances is emphasized.

**Table 1.** Measured non-dimensional coupling parameters. The dimensionless coupling parameter $\gamma_j$ was measured by exploiting relationships of the type (22) specialized to the resonant shunt of the $j$th piezoelectric element.

| $\gamma_1$ | $\gamma_2$ | $\gamma_3$ | $\gamma_4$ | $\gamma_5$ |
|---|---|---|---|---|
| $122 \times 10^{-3}$ | $9.54 \times 10^{-3}$ | $5.77 \times 10^{-3}$ | $2.98 \times 10^{-3}$ | $0.083 \times 10^{-3}$ |

**Table 2.** Measured piezoelectric capacitances. The capacitance $c_j$ (in nF) was evaluated in an experimental set-up where the $j$th piezoelectric is shunted to an inductor, by measuring of the inductance for which the electric circuit is tuned to a given mechanical resonance frequency.

| $c_1$ | $c_2$ | $c_3$ | $c_4$ | $c_5$ |
|---|---|---|---|---|
| 51.30 | 53.75 | 53.36 | 52.92 | 52.90 |

In figure 2 we show the plot of $\gamma(\alpha_0, \alpha_n)$ corresponding to the experimentally measured values of $\gamma_j$ and $c_j$ reported in tables 1 and 2 for the tested cantilever beam on which five piezoelectric transducers were positioned. In this case the optimal values of the boundary impedances are

$$\begin{aligned} \alpha_0^{\mathrm{opt}} &= 1 \Rightarrow R_0^{\mathrm{opt}}, \qquad L_0^{\mathrm{opt}} \to \infty, \\ \alpha_n^{\mathrm{opt}} &= -1 \Rightarrow R_n^{\mathrm{opt}}, \qquad L_n^{\mathrm{opt}} \to 0. \end{aligned} \tag{25}$$

They provide the optimal electric modal shape $\{\tilde{\psi}_j^{(1)^{\mathrm{opt}}}\}_{j=1}^n = \{\tilde{\psi}_j^{(1)}(1, -1)\}_{j=1}^n$ and the corresponding values of $\lambda_1^{\mathrm{opt}} = \lambda_1|_{(\alpha_0,\alpha_n)=(1,-1)}$ and $\gamma^{\mathrm{opt}} = \gamma(1, -1)$.

By applying theoretical results about lattices (see e.g. [10]), it can be shown that, for all identical piezoelectric transducers, results similar to (25) hold for a generic number $n$. Furthermore, it can be shown that $\lambda_1^{\mathrm{opt}}$ decreases monotonically with $n$ and, for $n \geqslant 2$, $\lambda_1^{\mathrm{opt}} \leqslant 1$. In addition, asymptotically (i.e. for large $n$)

$$\lambda_1^{\mathrm{opt}} \simeq \left(\frac{\pi}{2n}\right)^2, \tag{26}$$

and, from (21),

$$\begin{aligned} L_{\mathrm{opt}} &\simeq \frac{(\pi/2)^2}{n^2(\omega_1)^2 c} = \frac{1}{n}\frac{(\pi/2)^2}{(\omega_1)^2 c_{\mathrm{tot}}}, \\ R_{\mathrm{opt}} &\simeq \frac{1}{n}\frac{(\pi/2)^2}{\omega_1\sqrt{2/3}\gamma_{\mathrm{opt}} c_{\mathrm{tot}}}, \end{aligned} \tag{27}$$

where $c_{\mathrm{tot}}$ is the total capacitance of the $n$ piezoelectric transducers ($c_{\mathrm{tot}} = nc$). Equation (27) shows that by increasing the number of transducers $n$ (and reducing the size of each element to keep the total amount of piezoelectric material constant) the optimal line inductance decreases. At the same time, the optimal resistance decreases without affecting the quality factor of the floating impedance due to the parallel connection of $L_{\mathrm{opt}}$ and $R_{\mathrm{opt}}$. Let us note that if each transducer is shunted with a dedicated $RL$ impedance the required inductance and resistance increase with $n$.

### 3.2. Resistive network

In some applications, when low damping is sufficient, adjacent piezoelectric transducers can be interconnected by a simple floating resistor. In this case, the host beam is coupled to a resistive–capacitive network, made up of inherent capacitances of the transducers and the floating resistors. The passive electric network is the electric parallel of a viscoelastic layer on the beam surfaces. The corresponding refined equations of motion are derived from (1) by setting $L \to \infty$ and the coupling between the first mechanical and electrical modes is described by (15) for $\beta = 0$. Furthermore, the corresponding mobility $H_R(\omega)$ is $H_{RL}(\omega)|_{\beta=0}$.

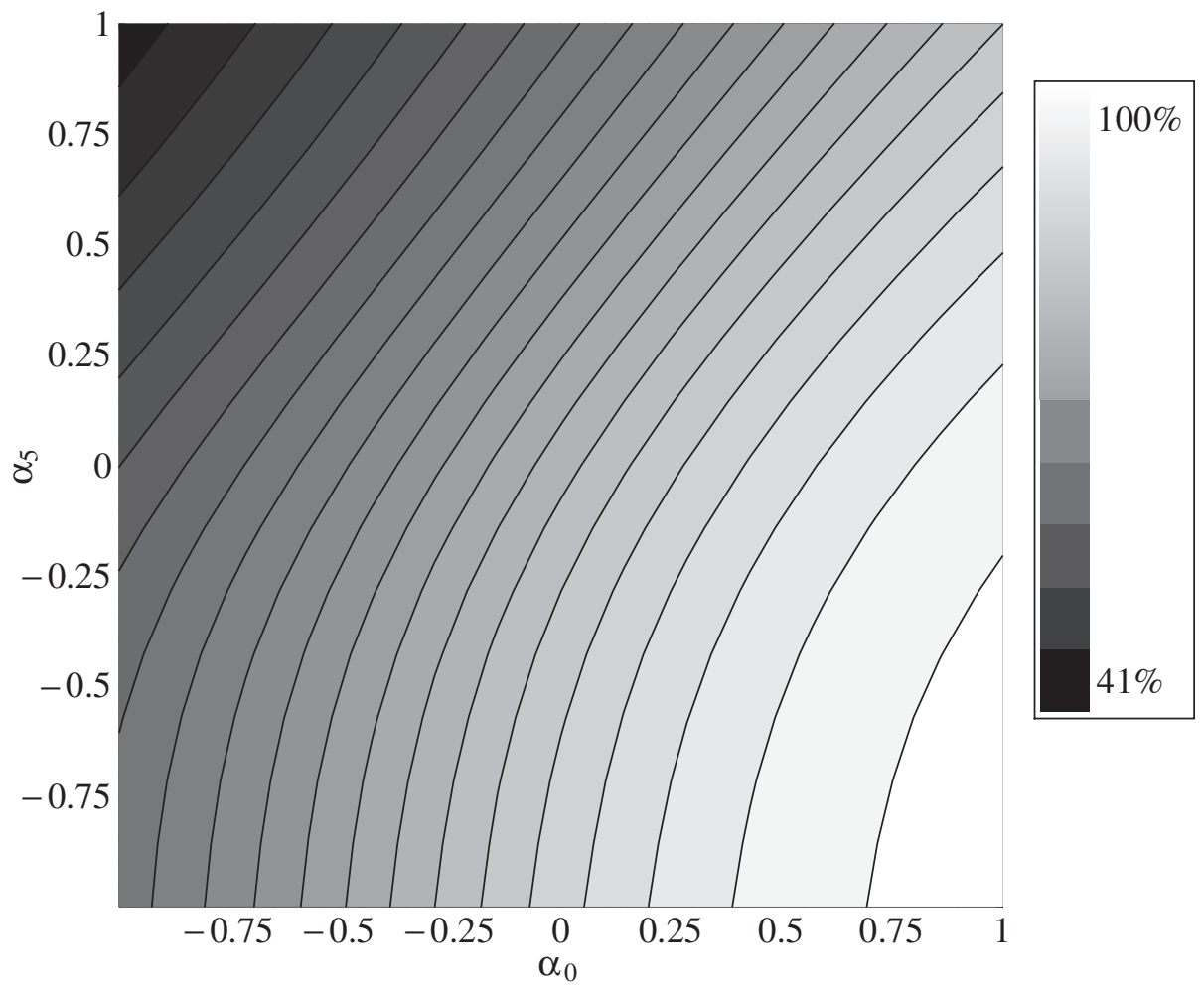

**Figure 2.** Contour plot of the coupling coefficient as a function of the boundary impedances (the values are scaled with respect to the maximum value 0.167).

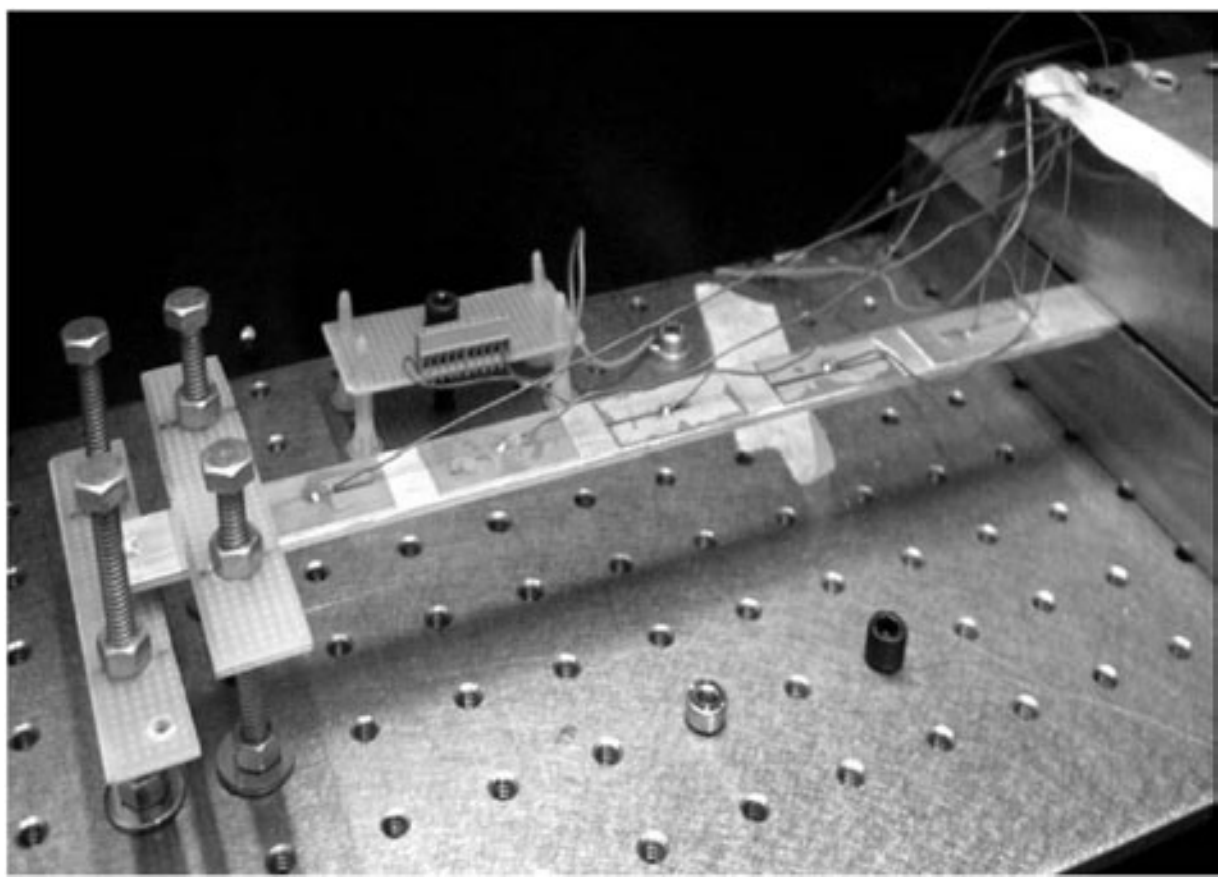

**Figure 3.** Picture of the PEM beam.

*3.2.1. Optimal damping coefficient and boundary conditions.* In this case the only parameter to be optimized is the damping coefficient $\delta$. The same $\mathcal{H}_\infty$ minimization criterion adopted for the $RL$ network yields:

$$\delta = \sqrt{\frac{8 + 10\gamma^2 + 3\gamma^4}{8 + 2\gamma^2}}. \tag{28}$$

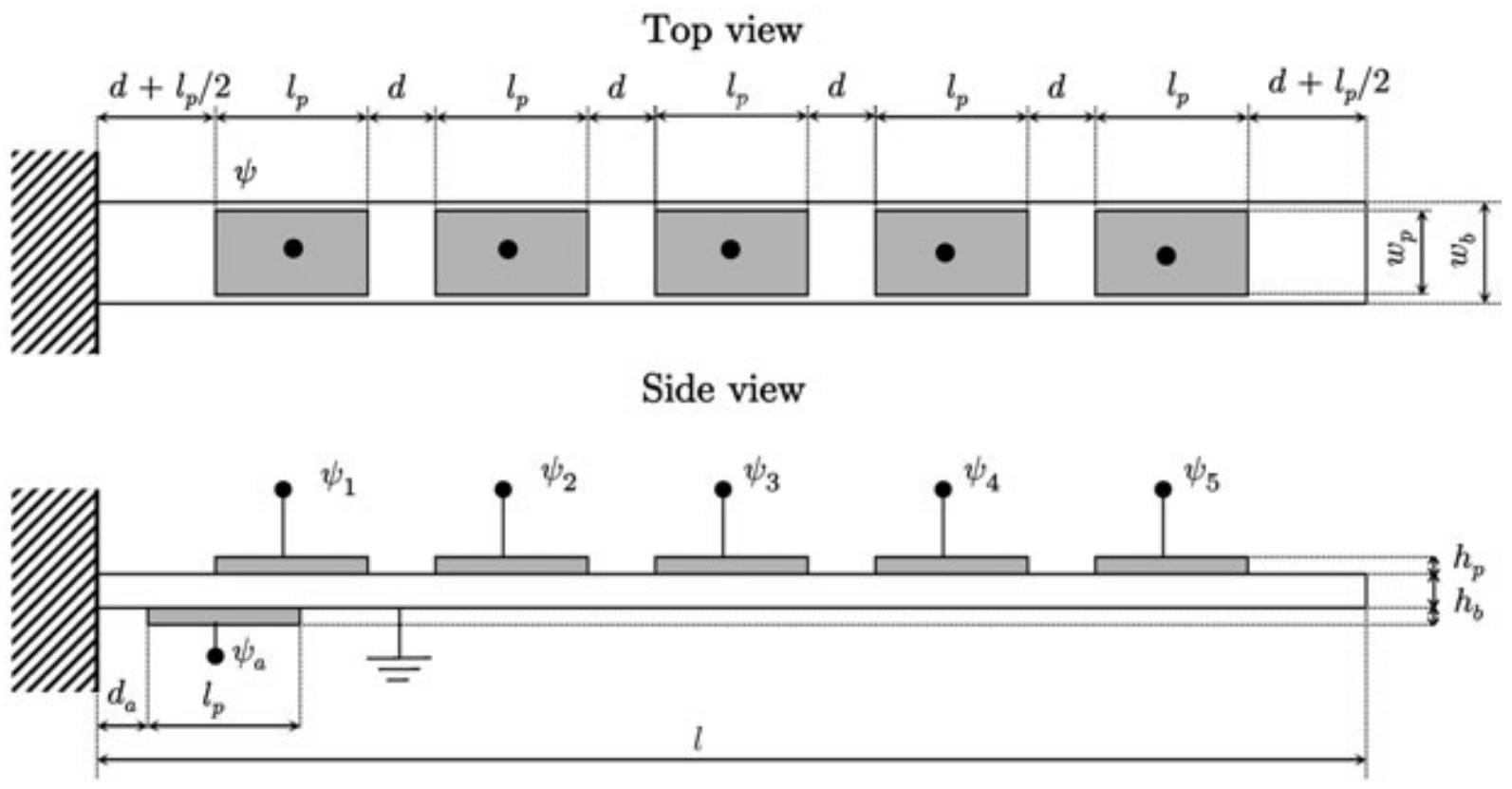

**Figure 4.** Geometry of the system.

Therefore, the optimal line resistance is found to be

$$R_{\text{opt}} = \frac{\lambda_1}{c}\frac{1}{\omega_1}\sqrt{\frac{8+2\gamma^2}{8+10\gamma^2+3\gamma^4}}. \tag{29}$$

This result can be analytically derived by noticing that there is a unique fixed point $F = (\omega_F, |H_R(\omega_F)|)$ with

$$\omega_F = \sqrt{1+\gamma^2/2}, \qquad |H_R(\omega_F)| = \sqrt{\frac{2(2+\gamma^2)}{\gamma^4}}; \tag{30}$$

and by putting

$$\|H_R\|_\infty = |H_R(\omega_F)|, \tag{31}$$

i.e.

$$\frac{\mathrm{d}}{\mathrm{d}\omega}|H_R(\omega)|\bigg|_{\omega=\omega_F} = 0. \tag{32}$$

Since in this case also the damping performance is increasing with the coupling coefficient $\gamma$ (see relation (30)), the optimal boundary resistances are defined by the dimensionless parameters $\alpha_0^{\text{opt}}$ and $\alpha_n^{\text{opt}}$ in (25).

A comparison between the damping effectiveness of the $RL$ network and the $R$ network, when the same boundary conditions are applied, is made from the ratio

$$\frac{\|H_R\|_\infty}{\|H_{RL}\|_\infty} = \sqrt{\frac{2}{\gamma^2}+1}, \tag{33}$$

which shows the higher performance of the $RL$ network, since the dimensionless coupling $\gamma$ is smaller than 1. Moreover, the smaller the piezoelectric coupling is, the higher the performance ratio (33) is. Hence, as a design guideline, when high piezoelectric coupling is available, purely resistive networks provide the relevant damping and passive controllers can get rid of inductances.

## 4. Experiments

### *4.1. Experimental set-up*

Experiments were conducted to validate the theoretical models presented above. An experimental set-up was built in order to test the corresponding damping effectiveness of both the $R$ network and the $RL$ network. Frequency response tests were performed on an aluminium (Al6061-T6) cantilever beam with five surface-bonded piezoelectric transducers made of PZT-5H piezoelectric ceramics (Piezo-System T110-H4E-602), illustrated in figure 3 and sketched in figure 4 (the corresponding geometrical properties are reported in table 3).

**Table 3.** Geometrical properties of the beam and PZT transducers (in mm).

| $l$ | $w_b$ | $h_b$ | $l_p$ | $w_p$ | $h_p$ | $d$ | $d_a$ |
|---|---|---|---|---|---|---|---|
| 273.6 | 19.5 | 1.90 | 35.6 | 17.8 | 0.27 | 10.0 | 5.0 |

**Table 4.** Nominal values of the electric components used to implement inductors.

| | |
|---|---|
| Deboo ($L = 130.5$ H) | $R = 2.7$ kΩ<br>$C = 17.9$ $\mu$F (polyester) |
| Antoniou ($L_5 = 19.01$ H) | $R_1 = 3$ kΩ<br>$R_2 = 1$ kΩ<br>$R_3 = 0$ kΩ<br>$R_4 = 1$ kΩ<br>$R_6 = 198$ Ω<br>$C_5 = 32$ $\mu$F (polyester) |

Piezoelectric transducers were bonded on the beam by a thin layer of non-conductive epoxy resin. Electric contact between the lower electrodes of the transducers and the grounded beam was achieved by applying a small spot of electrically conductive adhesive to the central region of the piezoelectric transducer, where interfacial stresses are low (see [13]). The frequency response of the beam was evaluated by applying, through an additional driving transducer, a sweep signal produced by a National Instruments AT-MIO-16E-10 D/A converter and by measuring the tip velocity with a Polytec OFV 350 laser velocimeter. The analogue laser output was converted by a National Instruments PCI-4452 A/D converter and a personal computer was used for digital signal processing and system identification.

The relevant measured parameters of the beam with the bonded piezoelectric transducers which appear in the modal model (15) are reported in tables 1 and 2. The first resonance frequency of the beam when all the piezoelectric elements are short-circuited is

$$\omega_m = 2\pi \times 20.44 \text{ Hz}. \tag{34}$$

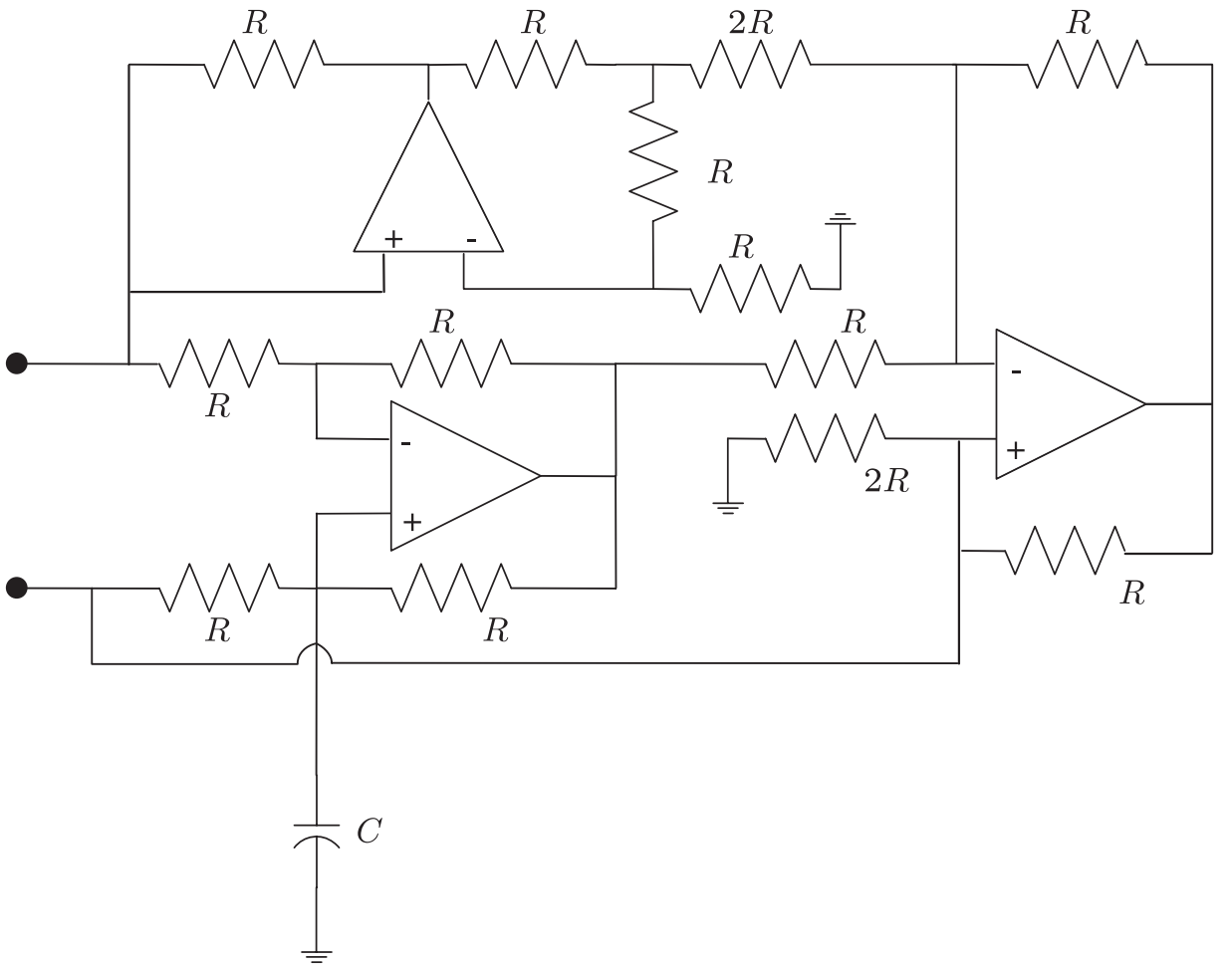

**Figure 5.** Schematics of the floating inductors.

From expressions (21) the optimal inductance and resistance for the $RL$ network are given by:

$$L_{\text{opt}} = 139.1 \text{ H}, \qquad R_{\text{opt}} = 123.2 \text{ k}\Omega. \tag{35}$$

The floating inductor can be simulated, according to [12], by exploiting the $RC$ circuit with three operational amplifiers depicted in figure 5. High-precision resistors must be used in order to reduce circuit losses, and guarantee the two-terminal behaviour of the simulated inductor. Careful choice of the dimensioning of the circuit components may yield very high quality factors, without affecting the maximum allowed voltage. Meanwhile, attention must be paid to undesired instability phenomena, which may eventually be compensated by introducing additional resistors connecting the circuit terminals to ground. The analysis of the introduced $RC$ circuit shows that the circuit is equivalent to a sole floating inductor, the inductance of which is

$$L = R^2 C. \tag{36}$$

From a practical point of view, inductance can be varied only by tuning the loading capacitance $C$, since otherwise simultaneous change of all the resistances is required.

In order to avoid simultaneous tuning of all the inductors, the internal resonance condition (21) is achieved by following a simpler approach. Indeed, by looking at the plots in figure 6 it is clear that variations of the terminal inductance $L_5$ in the range $[0, L]$ (i.e. $\alpha_5 \in [-1, 0]$) do not noticeably affect the electromechanical modal coupling $\gamma$, but, on the other hand, they do influence the electric resonance frequency. Therefore, the electric network can be tuned to the structural modal frequency by changing only the boundary inductance $L_5$.

A modified Antoniou circuit [11] (see figure 7) was used to simulate the tuning grounded inductor $L_5$. The corresponding equivalent inductance is given by

$$L_5 = \frac{R_1 R_4 R_6}{R_2} C_5. \tag{37}$$

Moreover, high quality factors can be achieved by varying the resistance $R_3$ (see [11]).

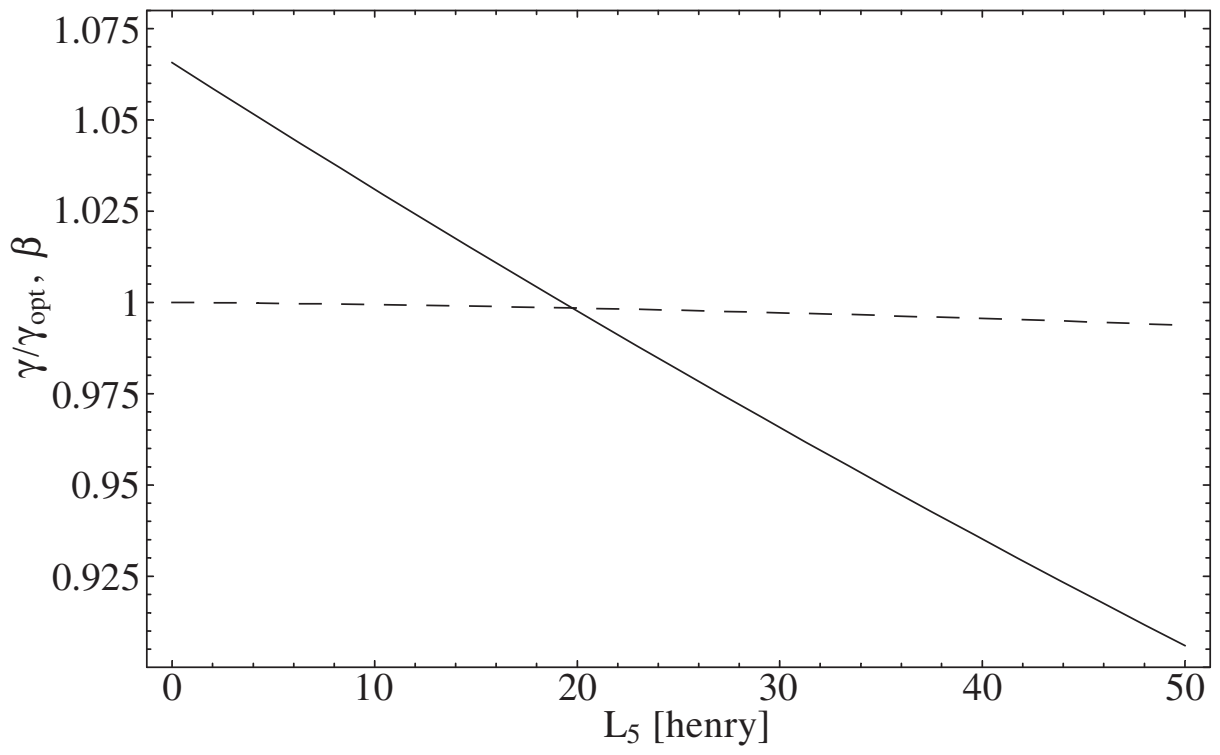

**Figure 6.** Tuning and coupling parameters $\beta$ (solid line) and $\gamma$ (dashed line) as functions of the terminal inductance when the line inductance is fixed to 130.5 H.

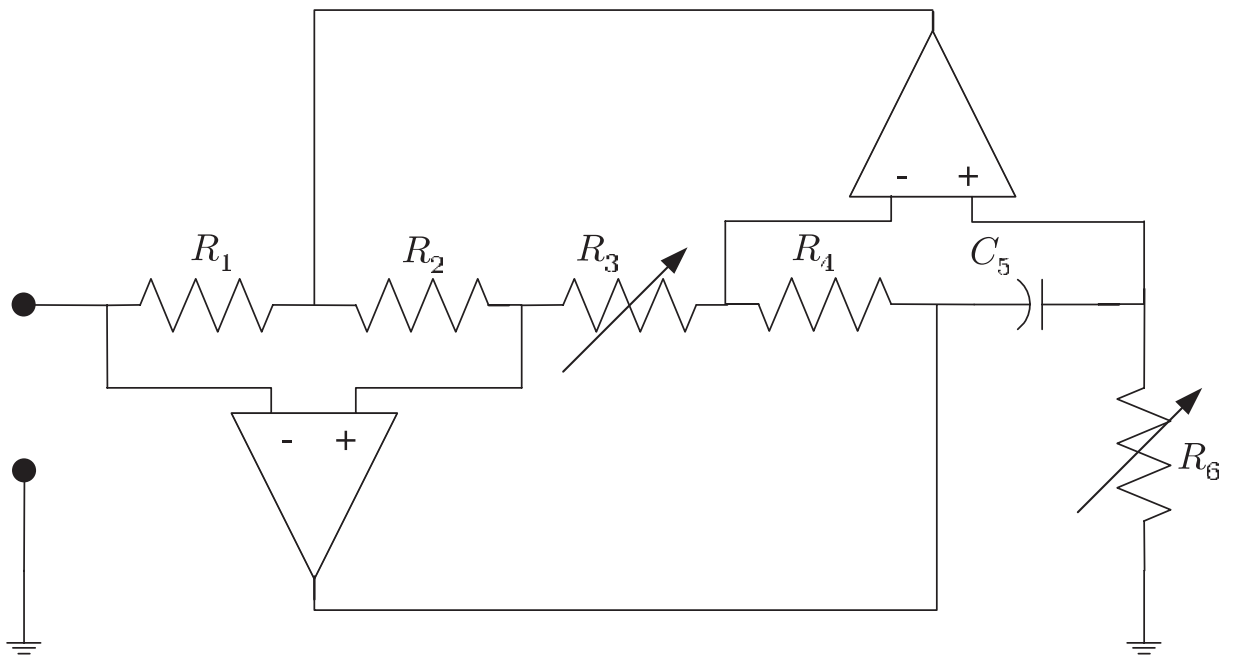

**Figure 7.** Schematics of the grounded inductor.

The values of the components employed for the realization of the four floating inductors and the grounded one are reported in table 4. High-voltage Burr–Brown OPA445AP FET-input operational amplifiers driven by a TTi EX752M dual-output power supply at ±30 V and high-precision resistors (±1%) have been used.

### *4.2. Results*

In this subsection, the theoretical predictions so far developed are validated through experiments. In particular, experimental mechanical mobility functions are shown for different values of the design electric parameters, resistances and inductances. Measurements in different frequency ranges are shown for both the $R$ and $RL$ network, optimized for damping the first structural mode.

*4.2.1. $RL$ network.* As described above, a $RL$ network tuned to the first mechanical mode has been obtained by using floating inductors of fixed inductance ($L = 130.5$ H) and an adjustable boundary inductor $L_5$, exploited for fine-tuning. By looking for the inductance $L_5$ for which condition (18) was satisfied, the optimal value $L_5 = 19.01$ H was experimentally found. In figure 8, we report corresponding experimental mobility functions for different values of the line resistances $R$ ($R_5$ was chosen to fulfil the proportional damping condition assumed by relations (3)). From equation (24), the corresponding coupling coefficient $\gamma$ is found to be almost equal to the maximum one achieved when $L_5$ vanishes, namely 0.167. From equation (22), the difference between the frequencies of the fixed points $S$ and $T$ can be evaluated to be 2.41 Hz, while from equation (21) the optimal resistance turns out to be 123 kΩ. These theoretical predictions, based on the reduced model described in section 2.2 and on the experimental measurements in tables 2 and 3, are in very close agreement with the experimental results implied by figure 8.

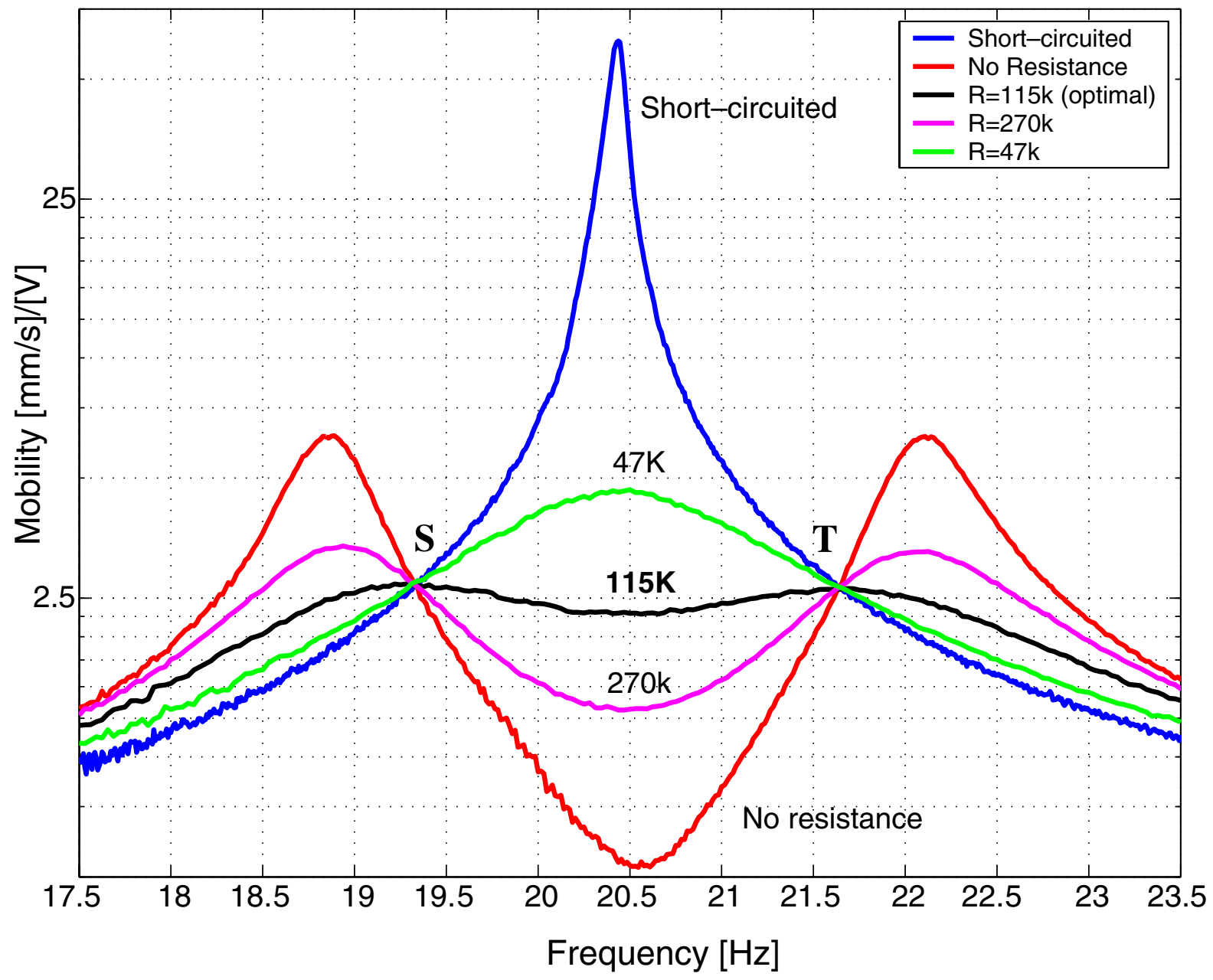

**Figure 8.** Experimental mechanical mobility around the first mode for the $RL$ network for different line resistances.

For higher modes the resonant coupling between the mechanical and electric subsystems is lost, and the $RL$ network essentially behaves as a purely resistive network [8]. Indeed, the proposed electric network allows for a resonant coupling only for a given beam mode since the remaining electric natural frequencies will generally be different from the structural ones. However, a non-negligible viscous-like damping is also added to higher modes, as shown by the experimental frequency responses in figures 9 and 10.

*4.2.2. R network.* The experimental implementation of the $R$ network is trivial, since only resistive elements interconnecting the electric terminals of adjacent transducers are required (according to (25) the last PZT is grounded). From equation (29), the optimal resistance is found to be

$$R_{\mathrm{opt}} = 17.6\ \mathrm{k}\Omega, \tag{38}$$

while, from (30), the fixed point frequency is 20.58 Hz. Furthermore, the performance ratio (33) between $R$ and $RL$ networks, for optimal boundary conditions, yields

$$\frac{\|H_R\|_\infty}{\|H_{RL}\|_\infty} = 8.53. \tag{39}$$

In figure 11, the frequency response for different values of shunting resistance are plotted, showing good agreement with theoretical results. For higher modes, the damping effectiveness is still relevant, as shown in figures 9 and 10. The experimental performance ratio (33) is 6.82; the discrepancy between the theoretical predictions and the experimental results can be put down to the neglect of the structural damping.

## 5. Conclusions

Devices for distributed passive electric damping of vibrations of structural beams have been presented. An array of uniformly distributed piezoelectric elements was positioned on a host beam and their terminals were interconnected by a passive electric network. Two different electric interconnections have been theoretically and experimentally examined:

(i) $R$ network (resistive connection of adjacent transducers);
(ii) $RL$ network (resistive–inductive connection of adjacent transducers).

The overall electromechanical system was treated as an Euler beam with material discontinuities and a lumped electric network, coupled together by electrically driven moment distribution and mechanically driven current sources. For design purposes, a reduced model with two degrees of freedom based on a single-mode approximation, for both the electric and mechanical subsystems, was introduced.

The electric networks were optimized with the aim of reducing the forced mechanical responses ($\mathcal{H}_\infty$ minimization on the mechanical mobility function). The optimal damping performance was shown to depend exclusively on the dimensionless piezoelectric coupling parameter between corresponding electrical and mechanical modes. Such piezoelectric modal coupling was maximized by posing an auxiliary optimization problem on the boundary conditions of the distributed electric network as a function of the mechanical modal shape to be damped.

The $RL$ network was shown to be an effective narrow-band controller, being the distributed version of a single resonant shunted piezoelectric transducer. The optimal line inductance of the distributed electric network was proved to be inversely proportional to the number of elements $n$ constituting the array of piezoelectric transducers. A similar dependence on $n$ was also found for the optimal line resistance, giving an idea of the technical feasibility of systems with a large number of transducers (the quality factor of $RL$ impedance connecting adjacent transducers results to be almost independent of $n$). Although the electric network was tuned to optimally damp the first mechanical mode, non-negligible damping was also added to higher modes (noting the results in [8], achieved by continuum modelling and wave propagation analysis). The $R$ network was shown to be a less effective controller at the first mode, while higher efficiency was shown at higher modes. Moreover, it was theoretically shown that the performance ratio between the $RL$ and $R$ networks depends uniquely on the effectiveness of the piezoelectric coupling. When high piezoelectric coupling is available the $R$ network ensures considerable multimodal damping and is preferable to the $RL$ network because of its simplicity. On the contrary, when the piezoelectric coupling is low the $RL$ network and resonant coupling are necessary for high performance.

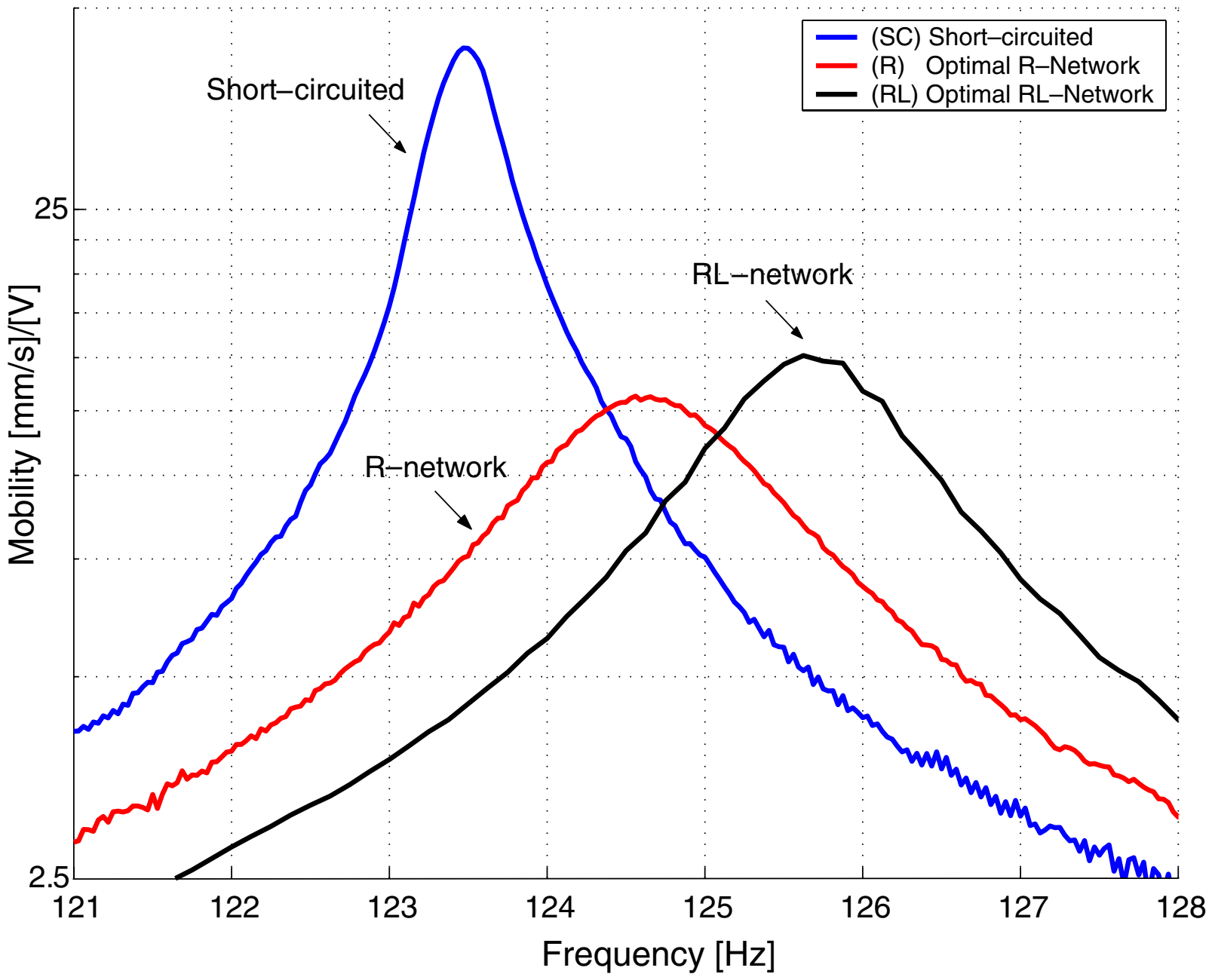

**Figure 9.** Experimental mechanical mobility around the second mode for optimal $R$ and $RL$ networks.

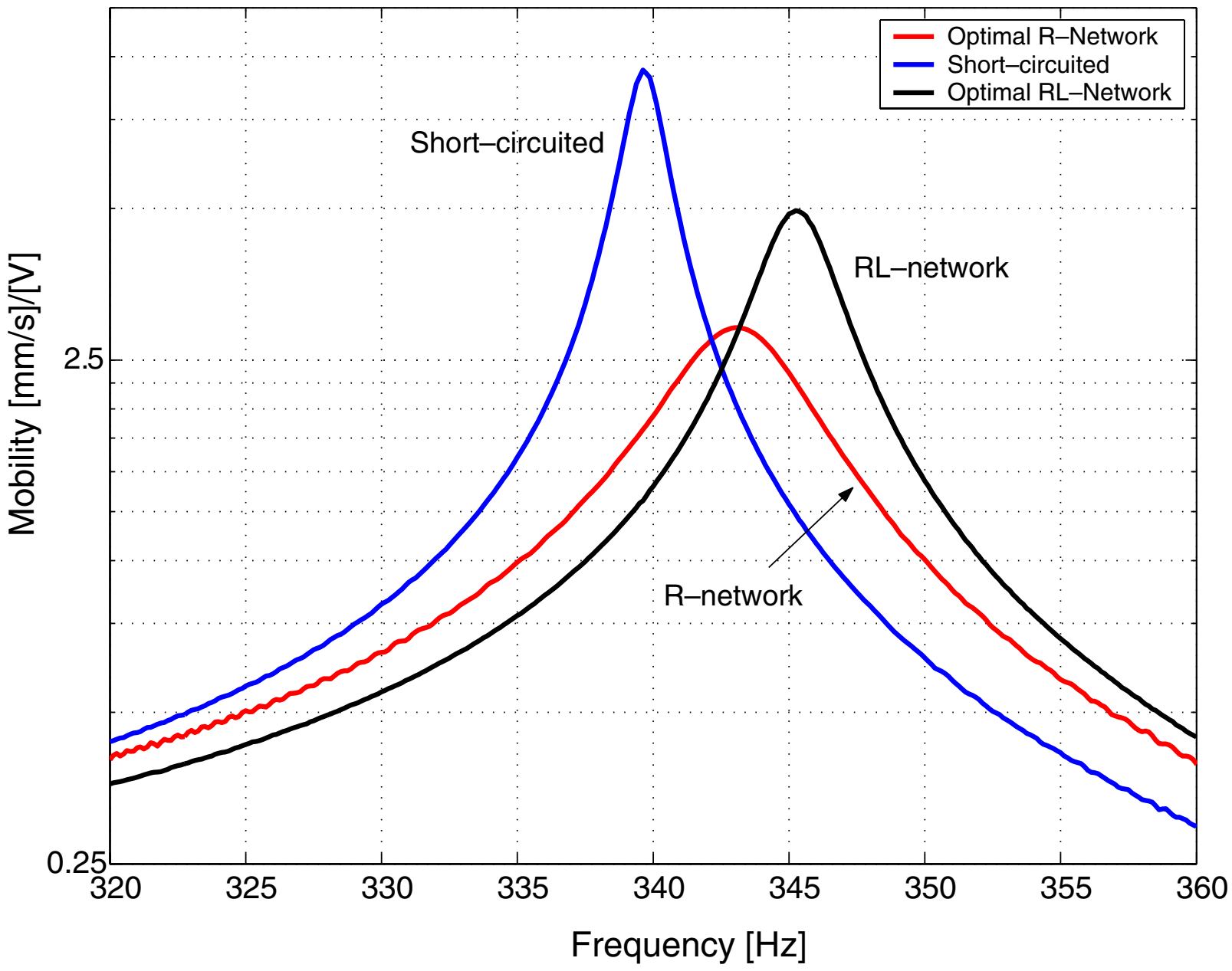

**Figure 10.** Experimental mechanical mobility around the third mode for optimal $R$ and $RL$ networks.

An experimental prototype was designed to assess the effectiveness of the proposed approach. Six piezoelectric ceramics were positioned on an aluminium cantilever beam, and one of them was used to excite the structure. Synthetic grounded and floating inductors with a high quality factor were realized. The maximum value of the mechanical mobility around the first mode is reduced by 95.8% by the $RL$ network and by 71.1% by the $R$ network. At the second and the third mode, vibrations are reduced by 65.9% and 47.1% by the $RL$ network and by 69.9% and 69.8% by the $R$ network.

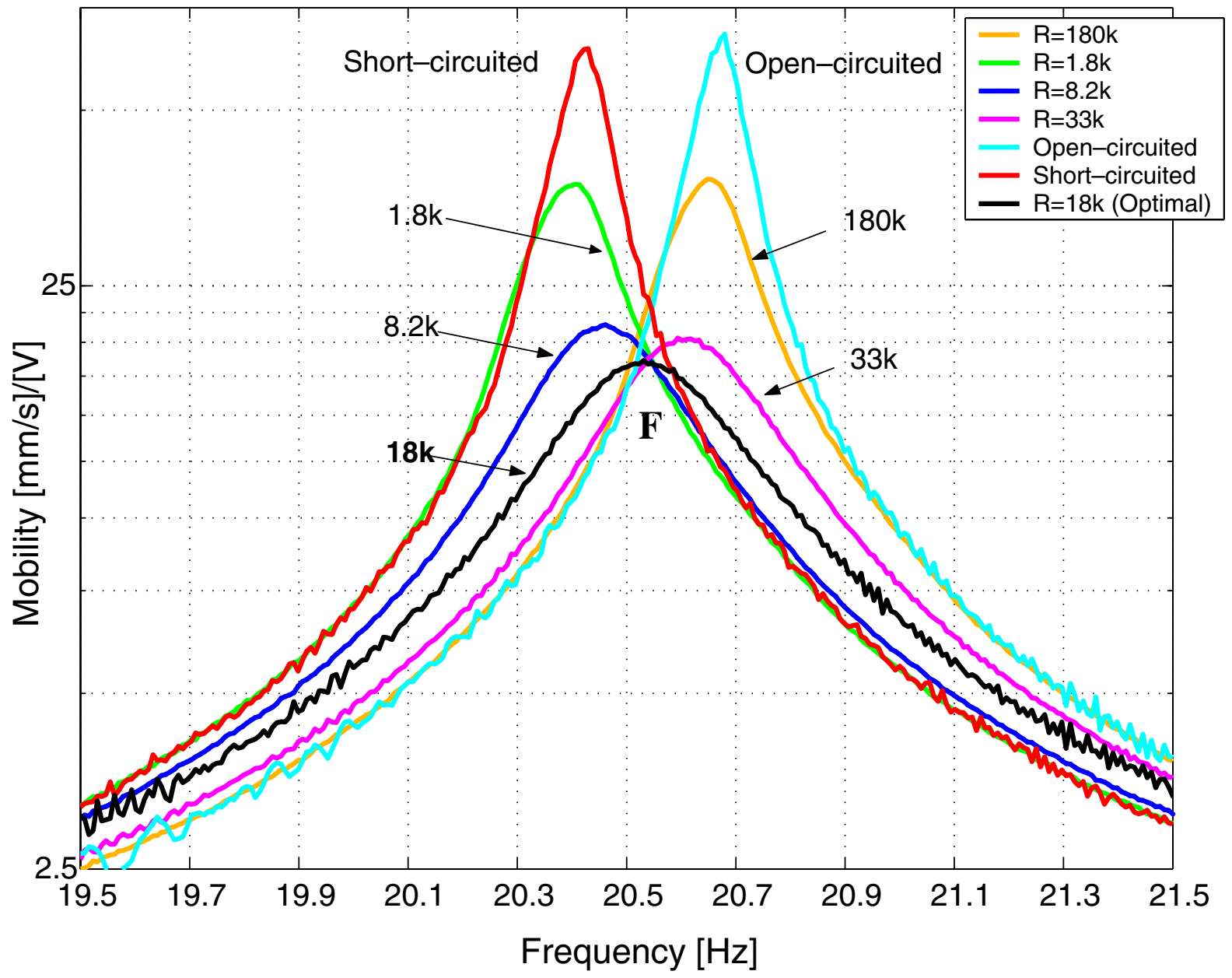

**Figure 11.** Experimental mechanical mobility around the first mode for the $R$ network with different line resistances.

The present work proves the effectiveness and technical feasibility of passive distributed control via piezoelectric transducers and interconnecting electric networks. The main advantages of the proposed approach lie in:

(i) multimodal damping,
(ii) reduction of the optimal inductance with respect to classical piezoelectric shunting,
(iii) decreasing optimal inductance for increasing number of piezoelectric elements.

A natural extension of this work will consist of experimental validation of the novel, more complex, interconnecting networks presented in [6]. These networks ensure *resonant* multimodal damping and further reduction of optimal line inductance. The necessary circuits contain floating inductors and transformers, and this work provides the theoretical and technological fundamentals for their actual implementation.

## Acknowledgments

The partial support of the Engineering Science and Mechanics Department of Virginia Polytechnic Institute and State University is gratefully acknowledged. This presented research has been also partially supported by MIUR, Ministero per l'Innovazione, l'Università e la Ricerca Fondi Ricerca PRIN 'Sintesi di circuiti piezoelettrici e tecniche di disaccoppiamento per il controllo di vibrazioni meccaniche' (protocollo 2001097882_003) and by the Università di Roma 'La Sapienza'. The authors would also like to thank Dott. Pier Mario Pollina for a detailed review of this work. Finally, the unconditional support of Professor Aldo Sestieri is acknowledged.